# Achieving Analytical and Cellular Control in Confined Spaces and Continuous Flow

*Sabeth Verpoorte - University of Groningen*

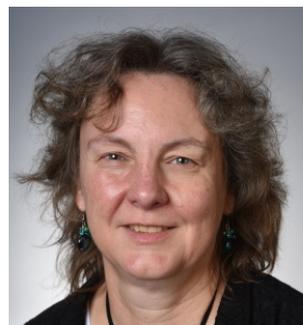

## Biography

E.M.J. (Sabeth) Verpoorte has more than 26 years of research experience in the microfluidics or lab-on-a-chip field. Her introduction to the field came in 1990, when she came from Canada as a postdoctoral researcher in the pioneering µTAS group headed by Professor A. Manz at Ciba Ltd., Basel, Switzerland. In July 1996 she became a team leader in the group of Professor Nico F. de Rooij at the Institute of Microtechnology (IMT), University of Neuchâtel, Switzerland, where her research interests concentrated on microfluidics for (bio)analytical applications. In 2003, Sabeth made a strategic switch to assume a Chair in the Groningen Research Institute of Pharmacy, making a foray into a new research environment dominated by cells, tissue and drug development. Ongoing projects involve the development of organ-on-a-chip systems to study drug metabolism (liver chip, gut chip), assess organ interactions (intestine-liver chip) and diagnose endothelial dysfunction. Efforts have also concentrated on continuous-flow particle separation strategies, paper microfluidics, as well as miniaturized analytical instrumentation (paper spray ionization, multidimensional chromatography). The acquisition of interdisciplinary projects involving scientists from the life sciences, chemical, and physics disciplines continues to be a driving force in her research. She is or has been involved in several international scientific organizations and journal editorial boards.

## Introduction

It has been over 25 years since the "miniaturized total (chemical) analysis system", or MicroTAS, concept was first presented by Andreas Manz at the Transducers'89 conference in Montreux, Switzerland. This concept, originating as it does from analytical chemists, is based on the idea that multiple sample handling steps could be integrated into interconnected microfluidic channel networks as the basis for "sample-in, answer-out" analytical devices. The MicroTAS thus embodies a strategy for automation of (bio)chemical processing, involving flow systems for sample transport, handling and analysis. The performance of these systems is enhanced by virtue of the fact that the solution volumes involved are small, from a few microliters down to amounts measurable in attoliters. The small dimensions of the MicroTAS are its main strength, as solution flows at this scale are well-defined and can be controlled with exquisite precision. It is not surprising, then, that the MicroTAS has evolved into microfluidics, the science and technology of systems for processing fluids in microchannels [1], and the more applications-oriented concept, lab-on-a-chip. Microfluidics technology offers researchers an unparalleled opportunity to create and control flow and force gradients in microenvironments. This unique ability can be further exploited to exercise dynamic control over the (bio)chemical environment in the confined space of a microfluidic channel by directed transport and generation of gradients on a molecular level. Though invented by analytical chemists, the potential of microfluidics has therefore been recognized and embraced by researchers in quite different fields for a host of applications other than analytical chemistry.

The MicroTAS has been successfully repurposed over the past twenty years or so for engineering cellular microenvironments which more faithfully mimic *in vivo* conditions for cell- and tissue-based studies. Organs-on-a-chip are a recent outgrowth of this effort, comprising systems that extend beyond microfluidic perfusion culture to include the possibility to recreate tissue structure, as well as intratissue transport and mechanical function. Successful examples have included devices for a "breathing" lung [2,3] and a "peristaltic" intestine [4]. With the ability to make better biomimetic organ-chips comes the desire, indeed necessity, to monitor cell growth and characterize cell behaviour under controlled conditions. Qualitative information about a culture can be obtained from cell appearance or tissue morphology. However, this information is often limited, and is better correlated with quantitative data for parameters such as pH, dissolved oxygen, nutrient consumption, release of biomarkers for viability and relevant metabolites to obtain a more complete picture of cell behavior. The possibility to integrate analytical functionality into organ-chips is now being exploited to actively control conditions in these biological systems. A major challenge for the analytical chemist has become how to probe miniscule culture volumes directly, with temporal and spatial resolution, using previously developed or newly invented microanalytical or lab-on-a-chip approaches.

This presentation will focus in part on our development of a microfluidic incubation system for drug metabolism studies using a "top-down" precision-cut tissue slice model. We have been able to leverage microfluidics to perform experiments which would not have been possible in a wellplate, including the demonstration of intestine-liver interaction *ex vivo*. The second part of the presentation will describe a more fundamental particle separation concept which relies on the generation of controlled bi-directional flow in microchannels having variable cross-sectional dimensions, by opposing pressure-driven and electrokinetic flows. While these two applications may seem quite disparate, they both represent a unique ability to perform experiments under flow with a degree of precision which would otherwise not have been possible without microfluidics. Moreover, continuing to improve our basic understanding of how to manipulate and control fluid, particle and molecular transport in confined microchannels is essential for fueling future microfluidic innovation in chemistry and the life sciences.

## A microfluidic multi-organ platform for drug testing and understanding disease

Early information on the metabolism and toxicity properties of new drug candidates is crucial for selecting the right candidates for further development. Preclinical studies rely on mammalian and human cell-based *in vitro* tests and animal studies to predict the *in vivo* behaviour of drug candidates in man, though neither are ideal predictors of drug behaviour in humans. While animal species sometimes share some similarity with humans with respect to drug metabolism, species differences remain unpredictable and substantial enough to make extrapolation of animal-derived data (both *in vivo* and *in vitro*) to humans often difficult. *In vitro* tests with cells of human origin, on the other hand, suffer from a different drawback, namely that cell monocultures lack the tissue structure and, importantly, proper expression of endogenous enzymes and transporters which dictate drug exposure and metabolism *in vivo*. Improving *in vitro* systems for preclinical studies thus remains a major challenge, motivated both by a desire to reduce the use of animals for practical and ethical reasons and improve the quality of data obtained with human cells. Microfluidics can be exploited to come closer to this goal, in combination with precision-cut liver slices (PCLS) as an improved, "top-down" organotypic system.

A major advantage of PCLS over hepatocytes or cell lines for *in vitro* studies on liver function is that all the different cell types are present in their natural tissue-matrix configuration. Clearly, however, human tissue is not readily available, and it is worthwhile considering how to perform a maximum number of informative experiments with small amounts of material.

We developed a novel microfluidic-based (biochip) system to perform metabolism and toxicity studies with mammalian PCLS under continuous flow (see Figure 1) [5,6]. A biochip was designed to integrate slices of liver in small, sequentially connected microchambers (25 µL) that can be continuously perfused. The requirements for this low-cost microfluidic device were biocompatibility, and the maintenance of constant oxygen concentration, pH, medium composition and temperature conditions over time. The resulting poly(dimethylsiloxane) (PDMS) flow-through microdevice integrated polycarbonate membranes to realize a well-defined medium flow, and integrated PDMS membranes to act as "breathing" membranes for oxygen supply. Perifusion of the slices suspended in solution (medium flowed around the slices) in the microchamber ensured a continuous influx of nutrients and oxygen and a continuous removal of waste products.

Studies with mammalian slices have shown that the use of microfluidics makes more sophisticated experimentation possible when compared to conventional well-plate formats. For example, it was possible to couple the outlets of three different microchambers containing perifused liver slices to a series of three HPLC injectors for automated, sequential on-line monitoring of metabolites. Because of the low volumes and flowrates involved, relatively high concentrations of metabolites could be achieved, enabling sensitive detection [7]. The biochip also enables the *in vitro* measurement of interorgan effects, by connecting the outlet of one chamber to the inlet of another chamber containing a slice from a different organ (Figure 1) [8]. The outlet of a microchamber containing a rat intestinal slice was coupled to the inlet of a microchamber containing a rat liver slice. The interplay between intestinal and liver slices was then assessed by introducing medium containing the primary bile acid, chenodeoxycholic acid. Upon exposure to this bile acid, the intestinal slice produced a protein, growth factor FGF15, which caused a down-regulation of the enzyme, cytochrome P450 7A1, responsible for bile acid synthesis in the liver, as known for the rat *in vivo*. By using our biochip system, we were able to show this organ interaction with *in vitro* testing. *In vivo* testing with animals is possible; however, *in vivo* testing of such organ interactions in humans is virtually impossible. The incorporation of human material in our device could bring the study of such human organ interactions *in vitro* closer to reality.

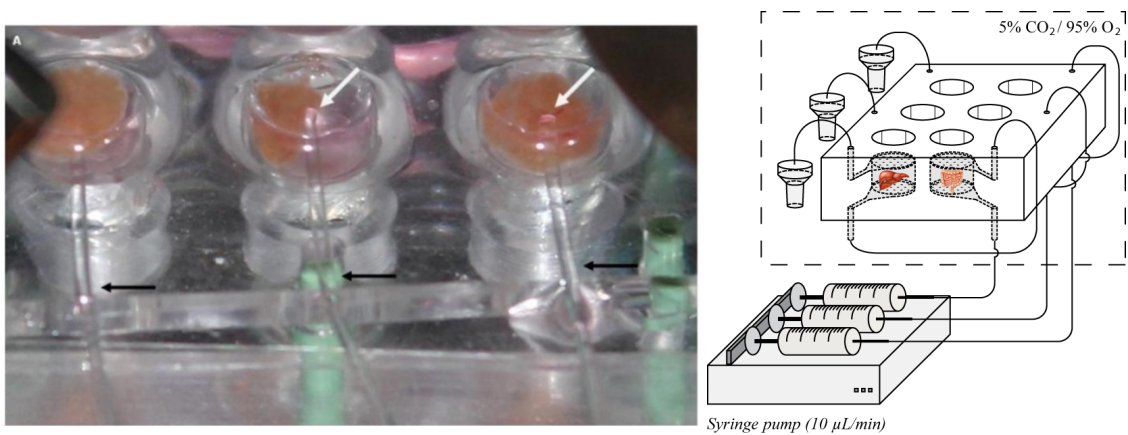

*Figure 1: (Left) Top view of a 10-layer poly(dimethylsiloxane) device with three 25-µL chambers, each containing a liver slice ( 4 mm diameter, 100-250 µm thick. The slices are suspended in medium flowing from the bottom of the device and exiting through the top. (Right) Biochip configuration for organ interaction experiment. An intestinal slice (rat) is located in the right chamber, the outlet of which is connected to the inlet of the second chamber containing a liver slice (also rat). The organs are represented figuratively.*

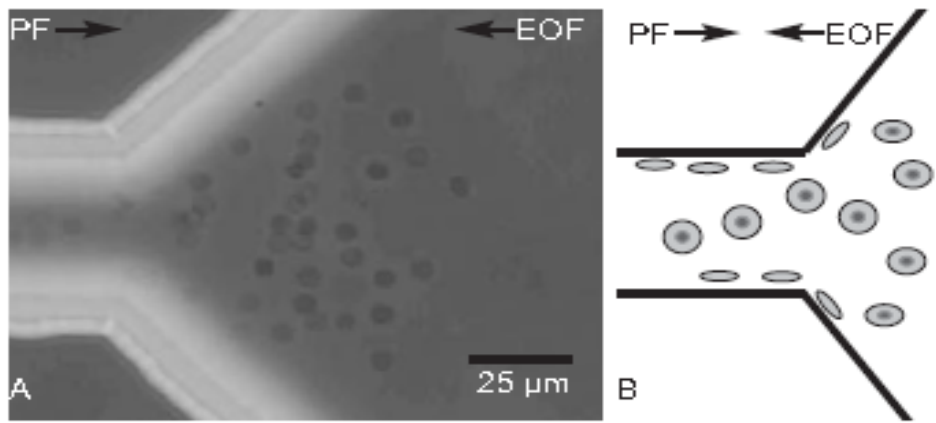

*Figure 2: Trapping of red blood cells (rat blood diluted with borate buffer). Note difference of cell alignment in different flow regions. E = 450 V, P = 4 mbar.*

## Two-dimensional orthogonal particle separations in nanoliter flows

As the range of applications for lab-chip technologies broadens, integrated methods for manipulating, sorting and separating particles in microfluidic systems become ever more relevant. Approaches based on laminar flows and applied electric fields are popular as they are easy to implement, provided careful thought has been given to microchannel geometry and layout. We have combined hydrodynamic and electrokinetic effects in a microchannel to preconcentrate and separate particles based on differences in their charge (zeta potential) [9]. This was done by opposing electro-osmotic and pressure-driven flow in a narrow microchannel to generate controlled bidirectional flow. Widening the channel at both ends leads to a recirculating flow pattern, in which polymer microspheres and biological particles (yeast, red blood cells, DNA) can be trapped and concentrated, a phenomenon we have termed flow-induced electrokinetic trapping (FIET). In fact, the approach depends on the relative ratio of applied electric field to pressure; tuning this ratio allows tuning of the trap for particles having a particular electrokinetic mobility [9]. Interestingly, FIET also allows separation of particles according to size, as particles falling into a narrow mobility range but having different sizes will circulate in the flow under a given set of conditions. Larger particles will occupy a smaller region of the channel and thus tend to drift towards the low-pressure end. In contrast, the net flux of smaller particles will be in the opposite direction with EOF, allowing these to escape the trapping channel [10]. To date, separations of binary mixtures of particles having either different charge or size have been demonstrated. In fact, the mechanisms for size-based and charge-based separation are very different, so that FIET enables orthogonal microparticle separations in nanoliter volumes.